\documentclass[onecolumn,fleqn,usenatbib]{mnras}

\usepackage[british]{babel}

\usepackage{amsbsy}
\usepackage{amssymb}
\usepackage{amsmath}
\usepackage{bm}
\usepackage{xfrac}
\usepackage{physics}

\usepackage{graphicx}

\hypersetup{
    bookmarksnumbered=true,
    breaklinks=false,
    citecolor=blue,
    colorlinks=true,
}

\usepackage[capitalise,compress]{cleveref}

\crefname{section}{Sec.}{Secs.}
\Crefname{section}{Section}{Sections}

\begin{document}


\title{The impact of nuclear reactions on the neutron-star g-mode spectrum}

\author[Counsell, Gittins \& Andersson] {A. R. Counsell, F. Gittins \& N. Andersson \\
Mathematical Sciences and STAG Research Centre, University of Southampton,. Southampton, UK}

\maketitle

\begin{abstract}
Mature neutron stars are expected exhibit gravity g-modes due to stratification caused by varying composition. These modes are affected by nuclear reactions, leading to complex (damped)  mode frequencies and the suppression of high order g-modes, in contrast with the common non-dissipative analysis which leads to an infinite g-mode spectrum. Focusing on the transition between the fast and slow reaction regimes, we examine the impact of nuclear reactions on the g-mode spectrum. The general framework for the analysis is presented along with sample numerical results for the BSk21 equation of state and the standard Urca reactions. 

\end{abstract}

\section{Introduction}

Neutron stars are highly compact objects that involve a rich and complex variety of physics. Due to this, there are (more or less) specific classes of oscillation modes associated with each aspect of the physics involved. One key feature is the varying composition of matter, introducing buoyancy as a restoring force in the equations of fluid dynamics \citep{Reisenegger}. This, in turn, leads to the presence of low-frequency gravity g-modes. The typical frequency of the leading neutron star g-mode is of the order of a few 100~Hz (depending on the matter equation of state) and the higher overtones lie at lower frequencies. As these modes rely on stable composition stratification for their existence, they are sensitive to nuclear reactions. Reactions will strive to reinstate beta equilibrium in the neutron star matter and hence lead to each g-mode becoming damped. The standard description is really only relevant in the limit of infinitely slow nuclear reactions. While this may apply to the highest frequency g-modes (the lowest overtones) there will always be modes that have low enough frequencies that the impact of finite reactions must be accounted for. Hence it is interesting---at least from a formal point of view---to consider the impact of nuclear reactions on the g-mode spectrum. This is the issue we will address in the following. The aim is to establish a more precise understanding of how the g-modes are affected by reactions and what happens to the mode spectrum when the reaction rate becomes fast compared to the dynamics.  This issue may not have immediate relevance for many astrophysical applications, but it connects with problems that involve the very high order g-modes. Two such problems immediately come to mind. First, the nonlinear saturation of modes that are driven unstable by gravitational-wave emission \citep{2001PhRvD..65b4001S}. Second, the so-called p-g instability which has been proposed to operate in neutron star binaries \citep{Weinberg_2013} and which involves the coupling of very high order pressure p-modes and g-modes to the dynamical tide.

The importance of stellar seismology had been appreciated since the work of \cite{Cowling} \citep[see][]{1980tsp..book.....C,1989nos..book.....U}. While originally the focus was on the mathematical formulation of the problem, recent X-ray timing observations and the  detection of gravitational waves from the  GW170817 neutron star merger event  have led to renewed focus on the problem \citep{GW170817,GW170817-2}. The vast majority of the literature on neutron star seismology has focused on oscillation modes that depend weakly on the precise nuclear physics below the crust, such as the fundamental f-mode or the pressure p-modes. Due to the complexity of the problem, different aspects of the physics have been studied one piece at a time. 

A considerable body of work has  also been dedicated to the neutron star g-modes \citep[see][to name a few]{Finn,McDermott,Miniutti}. This is natural, as these modes should be present in both mature and hot young neutron stars \citep[noting recent evidence that the g-modes may be excited during the proto-neutron star stage following a core-collapse supernova;][]{Burrows}. In general, the g-modes depend  on both the internal matter composition and the state of matter (e.g. the presence of neutron superfluidity). Unlike other oscillation modes such as the  f- and p-modes, the g-modes rely specifically on the conditions in the outer core of a neutron star, below the elastic crust. It is therefore of interest to ask what constraints on the nuclear physics could be made from stellar observations of g-modes. Specifically, for a mature neutron star the mode frequencies depend on the variation of the proton fraction with density, in turn subject to nuclear parameters like the nuclear symmetry energy. An observation of a specific g-mode frequency---perhaps as a tidal resonance induced during binary inspiral \citep{Wynn,Wynn2}---might then provide some insight into the nuclear physics and the equation of state (beyond bulk properties like the mass and radius of the star).


Building on previous proof-of-principle work by \citet{10.1093/mnras/stz2449}, this paper considers the spectrum of g-modes in a cold mature neutron star and the impact nuclear reactions have on the damping of oscillations. Instead of focusing on either the fast or slow reaction regimes, we consider how the modes behave for arbitrary (parameterised) reaction rates and examine the effect damping has on the crossover from slow to fast reactions. This provides a better understanding of the phenomenology of the problem and a clearer idea of what happens to the very high order g-modes in a realistic neutron star model. Our demonstrations are based on a specific equation of state, BSk21 \citep{BSk,BSk2}, which allows us to employ a  realistic description of the matter stratification.

The layout of the paper is as follows: in Section \ref{Perturbation}, we outline the background equations and physics that go into the problem. Section \ref{Plane-wave} examines the fast and slow reaction regimes using a plane-wave approach. Section \ref{Numerical} then solves the mode-problem for a general reaction rate, presenting the dimensionless equations and results. Finally, Section \ref{Conclusions} summarises the work and presents ideas for future continuation of this effort.

\section{The perturbation problem}\label{Perturbation}

Even though we are ultimately aiming for realism, here we will explore the impact of nuclear reactions on the g-modes of a stratified neutron star in the context of Newtonian gravity. This makes the results somewhat phenomenological, but as we will have to vastly exaggerate the reaction rates to demonstrate the features we are interested in this lack of precision is not a great concern. Given this context, we consider the problem for non-rotating stars, assuming that the oscillation modes---with label $n$ and frequency $\omega_n$---are associated with a polar perturbation displacement vector (expressed in the coordinate basis associated with the spherical polar coordinates $[r,\theta,\varphi]$)
\begin{equation}
    \xi^i_n(t,r,\theta,\varphi) =  \xi^i (r,\theta,\varphi) e^{i\omega_n t} \ ,
\end{equation}
with 
\begin{equation}
  \xi^i =  { 1 \over r} W_l Y_l^m \delta^i_r + { 1 \over r^2}  V_l \partial_\theta Y_l^m \delta_\theta^i + { i m \over r^2 \sin^2 \theta}    V_l Y_l^m  \delta_\varphi^i  \ ,
 \label{multipolesum}
\end{equation}
where the multipole amplitudes, $W_l$ and $V_l$, are functions of $r$ only and $Y_l^m(\theta,\varphi)$ are the usual spherical harmonics. Along with this, all scalar perturbations are expanded in spherical harmonics. That is, using   $\delta$ to indicate an Eulerian perturbation, we have the perturbed mass density
\begin{equation}
    \delta \rho_n = \delta  \rho (r,\theta, \varphi)  e^{i\omega_n t},
\end{equation}
with 
\begin{equation}
\delta  \rho =  \delta \rho_l Y_l^m,
\end{equation}
and similar for all other scalar quantities. In the following, whenever $p,\rho, \Phi$ are used without $\delta$ they refer to the value of the pressure, density and gravitational potential, respectively, of the background equilibrium star. As we are ignoring rotation, the background configuration is spherical so all associated quantities are only functions of $r$. 

Turning to the perturbed Euler equation,  we have 
\begin{equation}
-\omega_n^2 \xi_i + \nabla_{i} \delta  \Phi + { 1 \over \rho} \nabla_{i}\delta  p - { 1 \over \rho^2} \delta \rho \nabla_{i} p
 = 0 \ .
\end{equation}
This leads to the radial Euler component, 
\begin{equation}
    \frac{\dd \delta  p_l}{\dd r} - \left({ \delta  \rho_l  \over \rho}  \right) \frac{\dd p}{\dd r} =  { \omega_n^2 \rho \over r} W_l  
- \rho\frac{\dd \delta  \Phi_l}{\dd r} \ , 
\label{box1}
\end{equation}
 and then from the $\varphi$ component of the Euler equation we get,
\begin{equation}
    \omega_n^2 V_l = \delta  \Phi_l  + \frac{\delta  p_l}{\rho} \ .
\label{omegaVl}
\end{equation}

We also need the perturbed continuity equation,
\begin{equation}
    \delta \rho_l = - {W_l \over r} \frac{\dd \rho}{\dd r} -  {\rho\over r^2} \left[  \frac{\dd \left(r W_l \right)}{\dd r} - l(l+1) V_l\right] \ .
    \label{contens}
\end{equation}
Combining the last two equations, we get
\begin{equation}
 \frac{\dd \left(r W_l \right)}{\dd r}+   {r  W_l  \over \rho}\frac{\dd \rho}{\dd r}   
= -
  {r^2 \over \rho} \delta \rho_l + { l(l+1) \over \omega_n^2} \left(\delta  \Phi_l  + \frac{\delta  p_l}{\rho} \right) \ .
  \label{box2}
\end{equation}
Lastly, we have the  perturbed Poisson equation,
\begin{equation}
    \nabla^2 \delta \Phi = 4\pi G \delta \rho \ .
    \label{potens}
\end{equation}

Let us now add nuclear reactions to the problem. As the moving fluid is no longer in equilibrium, we need to consider additional parameters in the (perturbed) equation of state. A natural option, which helps  account for nuclear reactions driven by the deviation from beta equilibrium, is to introduce  the new variable $\beta = \mu_\mathrm{n}-\mu_\mathrm{p}-\mu_\mathrm{e}$ depending on the chemical potentials for  neutrons, protons and electrons (labelled n, p and e). In (cold) equilibrium, we then have $\beta=0$. This condition allows us to solve for the matter composition, e.g. the proton fraction $x_\mathrm{p}$ for a given density. For simplicity, we  assume pure npe matter and that the star is cold enough to be transparent to neutrinos, thus the relevant reactions will be the Urca reactions. As a consequence of this addition, the general equation of state will be a two parameter function $p=p(\rho,x_\mathrm{p})$ where $x_\mathrm{p}$ is the proton fraction. From \cite{10.1093/mnras/stz2449} we have,
\begin{equation}  
\Delta \beta = {\mathcal B \over 1 + i \mathcal A/\omega_n} \Delta \rho
\label{dbeta1} \ ,
\end{equation}
where $\Delta$ is the Lagrangian perturbation,
\begin{equation}
\mathcal B = \left( {\partial \beta \over \partial \rho}\right)_{x_\mathrm{p}} \ , 
\end{equation}
and
\begin{equation}
    \mathcal A =-\frac{1}{t_R} \ , 
\end{equation}
where $t_R$ is the characteristic reaction time (and the minus sign is a convention). Note that this means that $\Delta \beta = 0$ in the limit of very fast reactions, when $t_R\to 0$. In the opposite limit, when reactions are slow, we have $t_R\to \infty$ and therefore $\mathcal A\to 0$ and the matter composition is frozen. Moreover, given that the unperturbed star is assumed to be in chemical equilibrium we have $\Delta\beta=\delta\beta$. Using this and combining \eqref{dbeta1} and \eqref{contens} we get,
\begin{equation}
    i \omega_n \delta \beta_l- \mathcal A \delta \beta_l = -i \omega_n {\rho \mathcal B  \over r^2} \left[ \frac{\dd \left(r W_l \right)}{\dd r} - l(l+1) V_l \right] \ .
\label{rwl}
\end{equation}
As it is common to work with the perturbed pressure, we rewrite this equation as
\begin{equation}
    i \omega_n \delta \beta_l- \mathcal A \delta \beta_l =  i \omega_n  \mathcal B  \left( \delta \rho_l + {W_l\over r } \frac{\dd \rho}{\dd r} \right)  \ , 
\end{equation}
and we also need
\begin{equation}
\delta p_l = \left( {\partial p \over \partial \rho} \right)_{\beta} \delta \rho_l +  \left( {\partial p \over \partial \beta} \right)_{\rho} \delta \beta_l \ , 
\end{equation}
leading to
\begin{equation}
\delta p_l = \left[ \left( {\partial p \over \partial \rho} \right)_{\beta} +  { 1 \over 1 + i\mathcal A/\omega_n } \left( {\partial \beta \over \partial \rho}\right)_{x_\mathrm{p}}  \left( {\partial p \over \partial \beta} \right)_{\rho} \right] \delta \rho_l +  { 1 \over 1 + i\mathcal A/\omega_n } \left( {\partial \beta \over \partial \rho}\right)_{x_\mathrm{p}}  \left( {\partial p \over \partial \beta} \right)_{\rho}  \left(  {W_l\over r } \frac{\dd \rho}{\dd r} \right) \ .
\end{equation}
At this point it make sense to use the thermodynamic relation,
\begin{equation}
  \left( {\partial p \over \partial \beta} \right)_{\rho} \left( {\partial \beta \over \partial \rho}\right)_{x_\mathrm{p}}  =  \left( {\partial p \over \partial \rho}\right)_{x_\mathrm{p}}  -   \left( {\partial p \over \partial \rho} \right)_{\beta} \ ,
\end{equation}
to get
\begin{equation}
\delta p_l = \left\{ 1  +  { 1 \over 1 + i\mathcal A/\omega_n } \left[  \left( {\partial p \over \partial \rho} \right)_{\beta}^{-1} \left( {\partial p \over \partial \rho}\right)_{x_\mathrm{p}} -1 \right] \right\} \left( {\partial p \over \partial \rho} \right)_{\beta} \delta \rho_l +  { 1 \over 1 + i\mathcal A/\omega_n } \left[   \left( {\partial p \over \partial \rho} \right)_{\beta}^{-1} \left( {\partial p \over \partial \rho}\right)_{x_\mathrm{p}} -1 \right]  \left( {\partial p \over \partial \rho} \right)_{\beta}  \left(  {W_l\over r } \frac{\dd \rho}{\dd r} \right) \ .
\end{equation}
Next we define the speed of sound in equilibrium and at fixed proton fraction, respectively;
\begin{equation}
    c_s^2 = \left( {\partial p \over \partial \rho} \right)_{\beta} \ ,
\end{equation}
\begin{equation}
   \mathcal C^2 = \left( {\partial p \over \partial \rho}\right)_{x_\mathrm{p}} \ .
\end{equation}
These quantities are related to the commonly used adiabatic indices via
\begin{equation}
    c_s^2 = {p\Gamma\over \rho} \ , \quad \mbox{and} \ \quad \mathcal C^2 = {p\Gamma_1\over \rho}\ .
\end{equation}
We also introduce
 the density scale height (which is convenient as we want to avoid involving an explicit stellar  model in the plane-wave analysis below)
\begin{equation}
    {1\over H} = {1\over \rho} {\dd \rho \over \dd r } .
\end{equation}

With these definitions, we have
\begin{equation}
\delta p_l = \left[ 1  +  { 1 \over 1 + i\mathcal A/\omega_n } \left( {\mathcal C^2 \over c_s^2} -1 \right) \right]c_s^2 \delta \rho_l 
+  { 1 \over 1 + i\mathcal A/\omega_n } \left( {\mathcal C^2 \over c_s^2} -1 \right) c_s^2  \left(  {\rho W_l\over r H }  \right) \ .
\end{equation}
When solving the equations numerically later on, we  will want to remove $\delta \rho_l$ so we need
\begin{equation}
\delta \rho_l =  \left[ 1  +  { 1 \over 1 + i\mathcal A/\omega_n } \left(  {\mathcal C^2 \over c_s^2} -1 \right) \right]^{-1} \left\{ {1\over c_s^2} \delta p_l -  { 1 \over 1 + i\mathcal A/\omega_n } \left( {\mathcal C^2 \over c_s^2} -1\right)  \left(  {\rho W_l\over r H }  \right) \right\} \ .
\label{drho}
\end{equation}
Finally, we define the Brunt-V\"ais\"al\"a frequency as
\begin{equation}
\mathcal N^2 =  g^2 \left( {1\over c_s^2} - {1\over \mathcal C^2} \right) = {c_s^4 \over H^2}  \left( {1\over c_s^2} - {1\over \mathcal C^2} \right) = - {c_s^4  \over H^2 \mathcal C^2} \left( 1 - {\mathcal C^2 \over c_s^2} \right)\ ,
\end{equation}
where the local gravitational acceleration is
\begin{equation}
    g=\frac{\dd \Phi}{\dd r} = - {1\over \rho} {\dd p \over \dd r} = - {c_s^2 \over H} \ .
    \label{gracc}
\end{equation}
This leads to 
\begin{equation}
  {\mathcal C^2 \over c_s^2} -1 =  {\mathcal N^2  H^2\mathcal C^2 \over c_s^4 }  \equiv  \bar{ \mathcal N}^2 \ , 
\end{equation}
which is the dimensionless Brunt-V\"ais\"al\"a frequency, and we have
\begin{equation}
\delta \rho_l =  \left[ 1  +  {\bar{\mathcal N}^2 \over 1 + i\mathcal A/\omega_n }  \right]^{-1} \left\{ {1\over c_s^2} \delta p_l -  { \bar{\mathcal N}^2 \over 1 + i\mathcal A/\omega_n }  \left(  {\rho W_l\over r H }  \right) \right\} \ .
\label{gmodedensity}
\end{equation}

Finally, rewriting \eqref{box1} and \eqref{box2} we get
\begin{equation}
    \frac{\dd \delta p_l}{\dd r} - {c_s^2 \over H}   \delta  \rho_l =  { \omega_n^2 \rho \over r} W_l  - \rho \frac{\dd \delta \Phi_l}{\dd r} \ , 
\label{gmodepressure}
\end{equation}
and
\begin{equation}
 \frac{\dd \left(r \rho W_l \right)}{\dd r} = -r^2  \delta \rho_l + { l(l+1) \over \omega_n^2} \left(\rho \delta  \Phi_l  + {\delta  p_l} \right) \ . 
  \label{gmodew}
\end{equation}
The last three equations are  the main equations that will be used to determine the neutron star g-modes.

\section{{Plane-wave analysis}}\label{Plane-wave}

In order to gain intuition and help explain the numerical results later, it is useful to consider a local plane-wave analysis. First, we introduce
\begin{equation}
    \bar W_l = {\rho W_l \over r},
\end{equation}
to get the, fairly concise, equations
\begin{equation}
    \frac{\dd \delta  p_l}{\dd r} - {c_s^2 \over H}   \delta  \rho_l =   \omega_n^2 \bar W_l  - \rho \frac{\dd \delta  \Phi_l}{\dd r} ,
\end{equation}
\begin{equation}
 \frac{\dd \left(r^2 \bar W_l \right)}{\dd r} = -r^2  \delta \rho_l + { l(l+1) \over \omega_n^2} \left(\rho \delta  \Phi_l  + {\delta  p_l} \right), 
\end{equation}
and
\begin{equation}
\delta \rho_l =  \left[ 1  +  {\bar{\mathcal N}^2 \over 1 + i\mathcal A/\omega_n }  \right]^{-1} \left\{ {1\over c_s^2} \delta p_l -  { \bar{\mathcal N}^2 \over 1 + i\mathcal A/\omega_n }  \left(  {\bar W_l\over  H }  \right) \right\} ,
\label{drhol}
\end{equation}
From these equations it is evident that the  problem will change when we consider finite timescale reactions. In the limit of no reactions, when $\mathcal A=0$, we are dealing with an eigenvalue problem for $\omega_n^2$, so we will always have two roots $\pm \omega_n$. When $\mathcal A\neq 0$ the eigenvalues become complex and the symmetry of the mode pairs is less obvious.

As  we are mainly interested in the qualitative behaviour at this point, we introduce the Cowling approximation (which is expected to be reasonably accurate for the g-modes); setting $\delta \Phi_l =0$. Then, using \eqref{drhol} to remove $\delta\rho_l$ from the problem we get
\begin{equation}
    \frac{\dd \delta  p_l}{\dd r} -  \left[ 1  + {\bar{\mathcal N}^2 \over 1 + i\mathcal A/\omega_n }  \right]^{-1}   \left( {\delta  p_l \over H} \right) =   \left\{ \omega_n^2 - {c_s^2\over H^2} \left[ 1  +  {\bar{\mathcal N}^2 \over 1 + i\mathcal A/\omega_n }  \right]^{-1}  { \bar{\mathcal N}^2 \over 1 + i\mathcal A/\omega_n } \right\} \bar W_l,
\end{equation}
and
\begin{equation}
 \frac{1}{r^2} \frac{\dd (r^2 \bar W_l )}{\dd r}  -    \left[ 1  +  {\bar{\mathcal N}^2 \over 1 + i\mathcal A/\omega_n }  \right]^{-1} { \bar{\mathcal N}^2 \over 1 + i\mathcal A/\omega_n }  \left(  {\bar W_l\over  H }  \right) =  \left\{ {\mathcal L_l^2 \over \omega_n^2} -\left[ 1  +  {\bar{\mathcal N}^2 \over 1 + i\mathcal A/\omega_n }  \right]^{-1}\right\} {\delta  p_l \over c_s^2},
\end{equation}
where the Lamb frequency is defined as
\begin{equation}
    \mathcal L_l^2 = {l(l+1) c_s^2 \over r^2} \ .
\end{equation}

In order to explore the nature of the waves we are interested in, we now adopt the plane-wave approach with
\begin{equation}
    \hat p = \delta p_l \ , \qquad \hat W = r^2 \bar W_l \ , \qquad \partial_r \to ik .
\end{equation}
This leads to
\begin{equation}
   \left( ik  
   - \left[ 1  +  {\bar{\mathcal N}^2 \over 1 + i\mathcal A/\omega_n }\right]^{-1}{1 \over H} \right) \hat p = \left(   \omega_n^2 - \left[ 1  +  {\bar{\mathcal N}^2 \over 1 + i\mathcal A/\omega_n }\right]^{-1}{ \bar{\mathcal N}^2 c_s^2 \over 1 + i\mathcal A/\omega_n }  {  1\over H^2 } \right) {\hat W \over r^2},
\end{equation}
\begin{equation}
 \left( ik  - \left[ 1  +  {\bar{\mathcal N}^2 \over 1 + i\mathcal A/\omega_n }\right]^{-1}{ \bar{\mathcal N}^2 \over 1 + i\mathcal A/\omega_n }  {  1 \over H } \right) {\hat W \over r^2}
 =   \left(  {\frac{\mathcal L_l^2}{\omega_n^2} - \left[ 1  +  {\bar{\mathcal N}^2 \over 1 + i\mathcal A/\omega_n }\right]^{-1} }  \right){\hat p  \over c_s^2}.
\end{equation}

Here we make two simplifying assumptions. First we focus on short-wavelength motion, such that $k|H|\gg1$. Secondly, we assume that $\bar {\mathcal N}^2\ll 1$ as appropriate for weakly stratified matter (eventually leading to the anticipated low-frequency g-modes). 

It is now easy to see how the expected barotropic result emerges in the $\mathcal N^2\to0$ limit.  
 For  fast reactions, we get 
\begin{equation}
   \left( ik 
   - {1 \over H} \right) \hat p \approx ik \hat p =    \omega_n^2  {\hat W \over r^2},
\end{equation}
\begin{equation}
 ik {\hat W \over r^2}
 = -    \left( 1- {\mathcal L_l^2  \over \omega_n^2 }  \right){\hat p  \over c_s^2},
\end{equation}
leading to the dispersion relation
\begin{equation}
   \omega_n^2 = c_s^2 k^2 +   \mathcal L_l^2 . 
\end{equation}
This solution represents sound waves---the pressure p-modes in the full mode calculation later. Higher overtone modes have shorter scales (=larger $k$) and therefore lie at higher frequencies. In a neutron star, we expect to find an infinite set of high-frequency p-modes.

In the opposite limit of slow reactions, we have
\begin{equation}
    ik   \hat p  = \left(  \omega_n^2  - { \bar{\mathcal N}^2 c_s^2 \over  H^2 } \right) {\hat W \over r^2},
\end{equation}
\begin{equation}
   ik  {\hat W \over r^2}
 = -    \left( 1 - {\mathcal L_l^2  \over \omega_n^2 }   \right)  {\hat p  \over c_s^2}.
\end{equation}
Now we instead arrive at 
\begin{equation}
k^2 c_s^2  =  \left(   \omega_n^2     - { \bar{\mathcal N}^2 c_s^2 \over  H^2 } \right) \left( 1 - {\mathcal L_l^2  \over \omega_n^2 }  \right). 
\end{equation}
This equation has two sets of roots. 
If it is also the case that (effectively focusing of dynamics slower than the sound waves)
\begin{equation}
    \omega_n^2 \ll \mathcal L_l^2,
\end{equation}
then
\begin{equation}
\omega_n^2 \equiv \omega_0^2 \approx  {\mathcal N}^2 { \mathcal C^2 \over  c_s^2 } \left( {k^2 c_s^2 +\mathcal L_l^2 \over \mathcal L_l^2}
 \right)^{-1} \approx {\mathcal N}^2 { \mathcal C^2 \over  c_s^2 } {l(l+1) \over k^2 r^2 + l(l+1)}.
 \label{gfreq}
 \end{equation}
In the opposite limit, when 
\begin{equation}
    \omega_n^2 \gg \mathcal L_l^2,
\end{equation}
it is easy to see that we retain the p-modes from the barotropic case. In essence, the introduction of the stratification has added a set of low-frequency modes to the spectrum. These are the g-modes. It is easy to see that, as the wavelength decreases (=larger $k$) the frequency decreases. This agrees with the results of \cite{1989nos..book.....U} where in beta equilibrium, the g-mode frequency was shown to tend to
\begin{equation}
    \omega_n^2\approx\frac{\mathcal N^2\mathcal L_l^2}{c_s^2k^2+\mathcal L_l^2}.
\end{equation}
In a neutron star, with stable stratification, we expect to find  an infinite set of undamped, low-frequency g-modes. This changes when we consider the nuclear reactions.

For finite reaction rates, we have 
(focussing on the $\omega_n^2\ll\mathcal L^2_l$ case) 
\begin{equation}
   ik   \hat p = \left(   \omega_n^2 - { \mathcal N^2  \over 1 + i\mathcal A/\omega_n }  {  \mathcal C^2\over c_s^2 } \right) {\hat W \over r^2},
\end{equation}
\begin{equation}
ik  {\hat W \over r^2}
 =  {\mathcal L_l^2  \over \omega_n^2 } {\hat p  \over c_s^2}.
\end{equation}
That is, 
\begin{equation}
    {k^2 c_s^2 + \mathcal L_l^2 \over \mathcal L_l^2} \approx {  t_R  \over \omega_n t_R - i}  {  \mathcal C^2\over c_s^2 } {\mathcal N^2 \over \omega_n},
\end{equation}
or
\begin{equation}
\omega_n \approx {  t_R  \over \omega_n t_R - i}  {\mathcal N}^2 { \mathcal C^2 \over  c_s^2 } \left( {k^2 c_s^2 +\mathcal L_l^2 \over \mathcal L_l^2}
 \right)^{-1} =  { t_R  \over \omega_n t_R - i} \omega_0^2.
 \end{equation}
As expected, the reactions lead to complex-frequency (damped) oscillations.  This is as it should be, given that the reactions lead to bulk viscosity which damps the fluid motion \cite{Schmitt}. It is, however, easy to see that we retain the previous (undamped!) results in the fast/slow reaction limits.

In the general case, solving for the frequency we have
\begin{equation}
     \omega_n \approx {1\over 2t_R} \left[ i \pm \left( 4\omega_0^2 t_R^2 -1 \right)^{1/2} \right].
        \label{FastGs}
\end{equation}
This is the key result, showing how we retain the undamped modes in the slow-reaction limit. Taylor expanding for large $\omega_0t_R$ we see that
\begin{equation}
     \omega_n \approx \pm \omega_0 + {i \over 2 t_R}.
     \label{modepair}
\end{equation}
That is, the two modes from the non-reactive problem are both damped and symmetric with respect to the imaginary axis. The numerically determined modes retain this symmetry. Similarly, 
  it is easy to see that both roots  become purely imaginary in the fast-reaction limit. The two roots limit to $\omega_n=0$ and $i/t_R$ (which means that $\omega_n\to i\infty$ as $t_R\to 0$), respectively. In fact, it is easy to see that there are no oscillatory modes below the critical reaction timescale $t_R=1/2\omega_0$. At this point, the pair of modes from \eqref{modepair} merge on the imaginary axis. Above the critical reaction time, they split again and one mode moves towards the origin while the other moves towards $+i\infty$ as $t_R$ decreases. In this regime the mode solutions represent pure diffusion. Above the critical timescale, for larger values of $t_R$, we have damped oscillations.
 The implications of this behaviour are---at least formally---important. While the classic analysis suggests that the g-mode spectrum is infinite, for a realistic neutron star model this cannot be so. The (potentially very) high overtones, for which $\omega_0 t_R$ is small, will be overdamped. This accords with the results from \cite{10.1093/mnras/stz2449}. 

 Finally, from \eqref{FastGs} we see that, in the regime where the modes are oscillatory, we have $\omega_n^2\approx \omega_0^2$. In essence, we expect the modes to move along a quarter circle, from the real axis (when $t_R=\infty$) to the imaginary axis (at the critical reaction time). This prediction is testable with numerical solutions and we now turn to that problem. 

\section{Numerical results}\label{Numerical}

\subsection{Dimensionless Form}

In order to solve the mode equations numerically, it is convenient to work with a dimensionless formulation \citep{1989nos..book.....U}. Relaxing the Cowling approximation, we define the following variables,
\begin{equation}
    y_1=\frac{W_l}{r^2},
\end{equation}
\begin{equation}
    y_2=\frac{1}{gr}\left( \frac{\delta p_l}{\rho}+\delta\Phi_l \right),
\end{equation}
\begin{equation}
    y_3=\frac{1}{gr} \delta\Phi_l,
\end{equation}
\begin{equation}
    y_4=\frac{1}{g}\frac{\dd \delta\Phi_l}{\dd r}.
\end{equation}
Inserting these definitions into \eqref{gmodedensity}--\eqref{gmodew} and the perturbed Poisson equation \ref{potens} we arrive at the coupled differential equations
\begin{equation}
    r\frac{\dd y_1}{dr} = \left( - \frac{\dd \ln \rho }{\dd \ln r} - 3 \right)y_1 + \frac{l(l+1)}{\tilde{\omega}^2_n c_1}y_2 -  \left[ 1  +  {\bar{\mathcal N}^2 \over 1 + i\mathcal A/\omega_n }  \right]^{-1} \left\{ {1\over c_s^2} gr(y_2-y_3) -  { \bar{\mathcal N}^2 \over 1 + i\mathcal A/\omega_n }  \left(  {r y_1\over  H }  \right) \right\} ,
\end{equation}
\begin{equation}
    r\frac{\dd y_2}{dr} = \tilde{\omega}^2_n c_1y_1 - \left[ 1 + \frac{\dd \ln (\rho g)}{\dd \ln r} \right] y_2 + \frac{\dd \ln \rho}{\dd \ln r}y_3 +\left[ 1  +  {\bar{\mathcal N}^2 \over 1 + i\mathcal A/\omega_n }  \right]^{-1} \left\{ {r\over H} (y_2-y_3) - { \bar{\mathcal N}^2 \over 1 + i\mathcal A/\omega_n } \frac{c_s^2}{H^2} \left(  {ry_1 \over g }  \right) \right\} , 
\end{equation}
\begin{equation}
    r \frac{\dd y_3}{\dd r}= (1-U) y_3 + y_4,
\end{equation}
and
\begin{equation}
    r\frac{\dd y_4}{\dd r}= l(l+1)y_3-U y_4 + 4\pi G \rho \left[ 1  +  {\bar{\mathcal N}^2 \over 1 + i\mathcal A/\omega_n }  \right]^{-1} \left\{ {1\over c_s^2}  r^2 (y_2-y_3) -  { \bar{\mathcal N}^2 \over 1 + i\mathcal A/\omega_n }  \left(  { r^2 y_1\over g H }  \right) \right\} .
\end{equation}
The equation for $y_3$ is obtained from the trivial relationship between $y_3$ and $y_4$ and  we have the used the same definitions as \cite{1989nos..book.....U} where
\begin{equation}
    \Tilde{\omega}^2_n = \frac{\omega^2_n}{GM/R^3},
\end{equation}
\begin{equation}
    c_1=\left(\frac{r}{R}\right)^3\frac{M}{m},
\end{equation}
\begin{equation}
    m=\int^{r}_{0}4\pi \rho r^2 \,dr\,
\end{equation}
\begin{equation}
    U= 2 + \frac{\dd \ln g}{\dd \ln r} = \frac{4\pi\rho r^3}{m} ,
\end{equation}
plus we define
\begin{equation}
    Z=-\frac{\dd \ln \rho}{\dd \ln r} = -\frac{r}{H},
\end{equation}
with $M$ and $R$ being the total mass and radius of the background star, respectively. 
Using \eqref{gracc}
we can rewrite the equations as
\begin{equation}
    x\frac{\dd y_1}{\dd x}=(Z-3)y_1+\frac{l(l+1)}{\Tilde{\omega}^2_n c_1}y_2-Z\left[ 1  +  {\bar{\mathcal N}^2 \over 1 + i\mathcal A/\omega_n }  \right]^{-1}\left\{y_2-y_3+{\bar{\mathcal N}^2 \over 1 + i\mathcal A/\omega_n }y_1 \right\},
\label{DimLess1}
\end{equation}
\begin{equation}
    x\frac{\dd y_2}{\dd x}=\Tilde{\omega}^2_n c_1 y_1 + (Z-U+1)y_2-Zy_3-Z\left[ 1  + {\bar{\mathcal N}^2 \over 1 + i\mathcal A/\omega_n }  \right]^{-1}\left\{y_2-y_3+{\bar{\mathcal N}^2 \over 1 + i\mathcal A/\omega_n }y_1 \right\},
\label{DimLess2}
\end{equation}
\begin{equation}
    x\frac{\dd y_3}{\dd x}=(1-U)y_3+y_4,
\label{DimLess3}
\end{equation}
\begin{equation}
    x\frac{\dd y_4}{\dd x}=l(l+1)y_3-Uy_4+UZ\left[ 1  +  {\bar{\mathcal N}^2 \over 1 + i\mathcal A/\omega_n }  \right]^{-1}\left\{y_2-y_3+{\bar{\mathcal N}^2 \over 1 + i\mathcal A/\omega_n }y_1 \right\},
\label{DimLess4}
\end{equation}
where
\begin{equation}
    x=\frac{r}{R},
\end{equation}
is a dimensionless radial coordinate.
This system of equation can be compared to that used by, for example, \cite{1989nos..book.....U} and it is easy to confirm that our set limits to the usual one when $\mathcal A\to 0$. 

Next we need four boundary conditions: two regularity conditions at the centre of the star and two conditions at the surface. Near the origin, as $r \rightarrow 0$, a Taylor expansion reveals that the we should have
\begin{equation}
    c_1\tilde{\omega}^2_n y_1-l y_2=0,
\label{BC1}
\end{equation}
\begin{equation}
    ly_3-y_4=0.
\label{BC2}
\end{equation}
Meanwhile, near the star's surface, as $r \rightarrow R$, we need to impose the vanishing of the Lagrangian pressure perturbation and the continuity of the gravitational potential and its derivative. These conditions take the form
\begin{equation}
    (l+1)y_3+y_4=0,
\label{BC3}
\end{equation}
and
\begin{equation}
    y_1-y_2+y_3=0.
\label{BC4}
\end{equation}
These relations follow by first requiring that $\delta\Phi_l$ vanishes at infinity and secondly that there exists a distinct boundary of the star where $\rho,p \approx 0$, thus $\Delta p=0$. These are the same boundary conditions as in the usual calculation \citep{1989nos..book.....U}. We can also see from this that our equations have no dependence on $m$ which is to be expected due to the spherical symmetry of the background star. Finally, to close the system we need an equation of state. We will discuss our chosen model below. 

\subsection{Results}

First of all, in order to gain confidence in the numerical implementation, we reproduce (and check) some results from the literature. The natural choice is to consider a $\Gamma=2$  $(n=1)$ polytrope with a constant $\Gamma_1$ representing the stratification \citep{Pnigouras_2015}. In particular,  in work by \cite{Scaling}, the numerical results show the following relations for the $l=2$ g-modes of a non-rotating Newtonian neutron star with a polytropic equation of state
\begin{equation}
    \tilde\omega_n \propto (\Gamma_1-\Gamma)^{1/2},
    \label{OmegaScale}
\end{equation}
\begin{equation}
    \tilde Q_n \propto \Gamma_1 - \Gamma,
    \label{QScale}
\end{equation}
where $\tilde Q_n$ is the dimensionless overlap integral defined as
\begin{equation}
    \tilde Q_n = \frac{1}{MR^l}\int_{0}^{R} \delta\rho_l  r^{l+2}\,dr , 
    \label{Overlap}
\end{equation}
also commonly referred to as the mass-multipole moment of each mode in the literature.

\begin{figure}
    \centering
    \includegraphics[width=10cm]{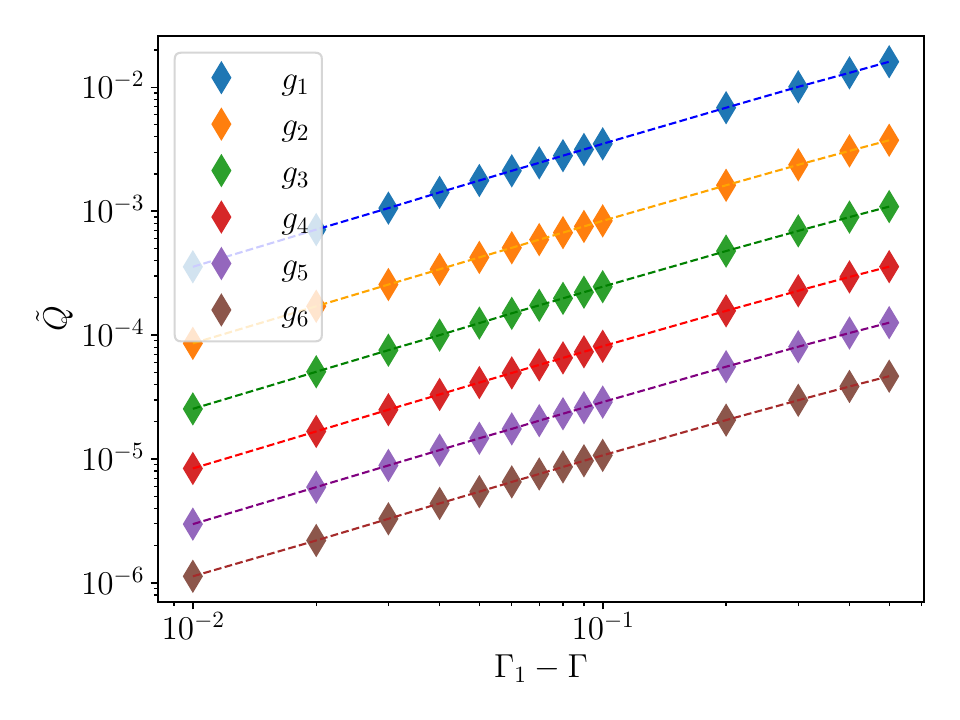}
    \includegraphics[width=10cm]{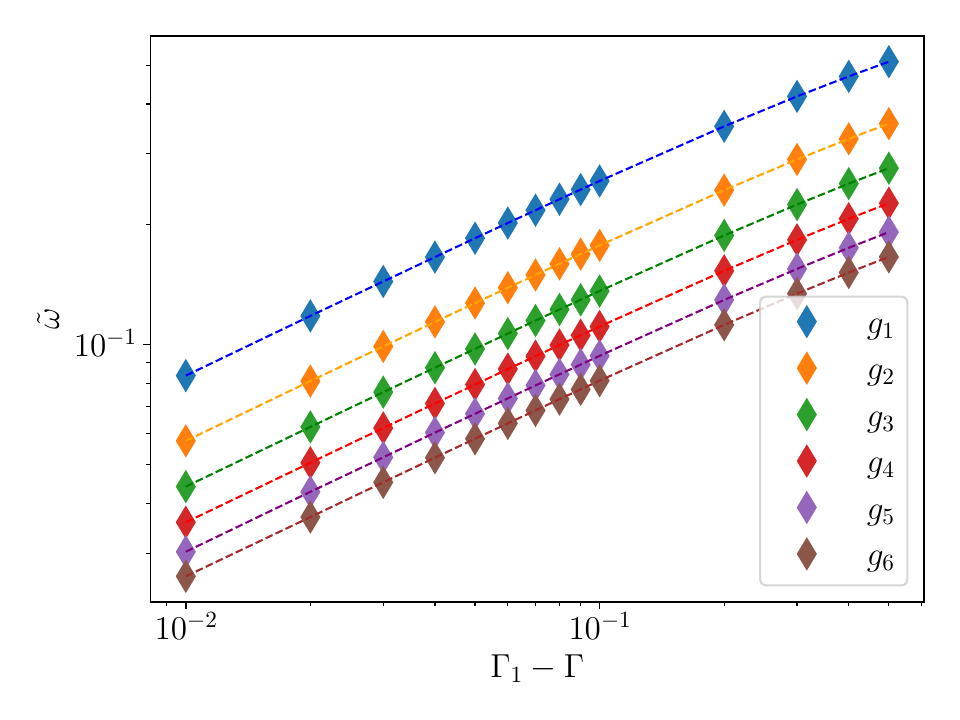}
    \caption{Plots of the dimensionless overlap integral $\tilde Q_n$ on the left and dimensionless frequencies $\tilde\omega$ on the right vs $\Gamma_1-\Gamma$ on log-log scales for the first 6 g-modes. The background star is polytropic with $\gamma=2$, $R=10$~km and $M=1.4M_\odot$. For the overlap integrals the slopes range from $0.95 - 0.98$ and for the frequencies the slopes range from $0.46-0.47$.}
    \label{Scales}
\end{figure}

Using the results from our code we obtain the results shown in Figure \ref{Scales} for the first 6 g-modes, where $\Gamma=2$ and $\Gamma_1$ is varied from $2.01-2.5$. This allows us to confirm the suggested scaling relations. Using a linear regression function, for the overlap integrals the slopes range from $0.95 - 0.98$ and for the frequencies the slopes range from $0.46-0.47$. This is in agreement with \eqref{OmegaScale}--\eqref{QScale} allowing for small numerical errors. 

\subsection{Realistic stratification}

In order to make the model more realistic we introduce stratification motivated by the BSk family of equations of state \citep{BSk,BSk2}. Specifically, mainly as proof of principle, we provide results for the BSk21 model \citep[see, also,][]{FabianRmodes,FabianRmodes2}. First, from BSk21 table, the relationship $p(n_\text{b})$ and $\bar{\mathcal N}^2 (n_\text{b})$ are calculated, where $n_\text{b}$ is the baryon number density. From this, $n_\text{b}$ is converted to mass density using $\rho=m_\text{b} n_\text{b}$, where $m_\text{b}$ is baryon mass. The pressure and density profiles as function of radius then follow (as usual) from the Newtonian stellar structure equations. Having fixed the equation of state, we still need to pick a sample neutron star. The particular model we consider has central density $\rho = 5.10 \times 10^{15} \mathrm{g\  cm}^{-3}$, which leads to a radius of $R = 13.49$~km and a mass of $M = 1.43 M_\odot$. Equations \eqref{DimLess1}--\eqref{DimLess4} along with the boundary conditions \eqref{BC1}--\eqref{BC4} were solved numerically using this background star and for different reaction rates by varying $\mathcal{A}$.

From the behaviour of the radial eigenfunction, $W_l(r)$, and the mode-frequency we determine if a mode is a p- or g-mode. Gravity modes tend to have small $\omega_n$ and a $W_l(r)$ that shows prominent features well beneath the surface of the star. Meanwhile, p-modes will have higher $\omega_n$ and a $W_l(r)$ that shows prominent features close to the surface. The fundamental f-mode can be thought of as the p-mode with the lowest frequency. One can also easily distinguish between overtones of the p- and g-modes. As the overtone $n$ increases, $\omega_n$ will decrease for g-modes but increase for p-modes (as per the plane-wave discussion). Also by examining $W_l(r)$ again, the number of nodes in the diagram is roughly equal to $n$, allowing us to identify the specific overtones.

\begin{figure}
    \centering
    \includegraphics[width=8cm]{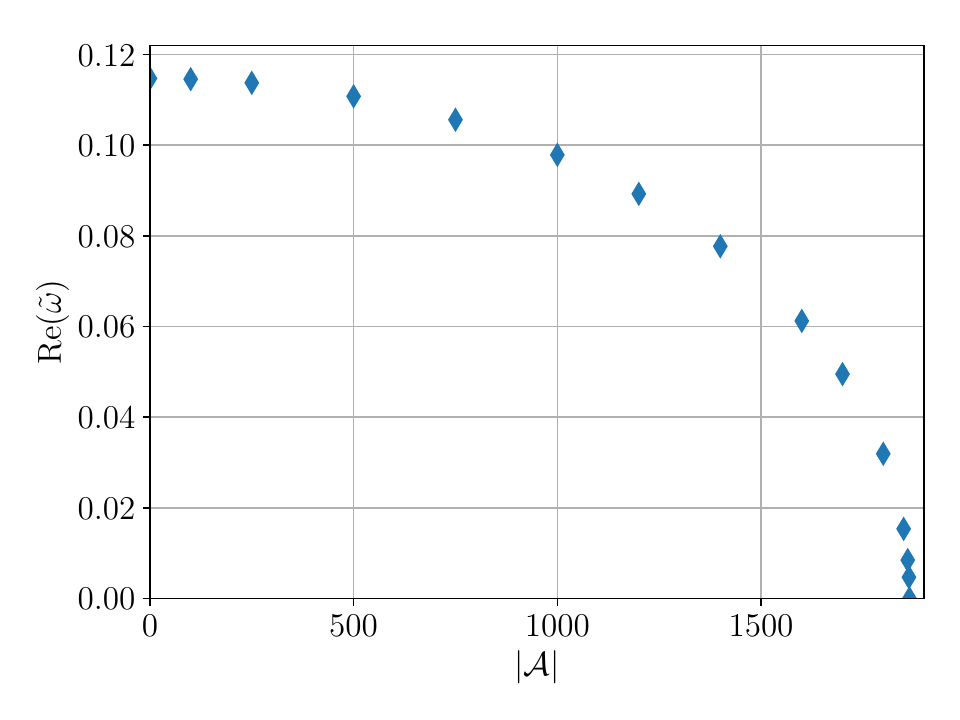}
    \includegraphics[width=8cm]{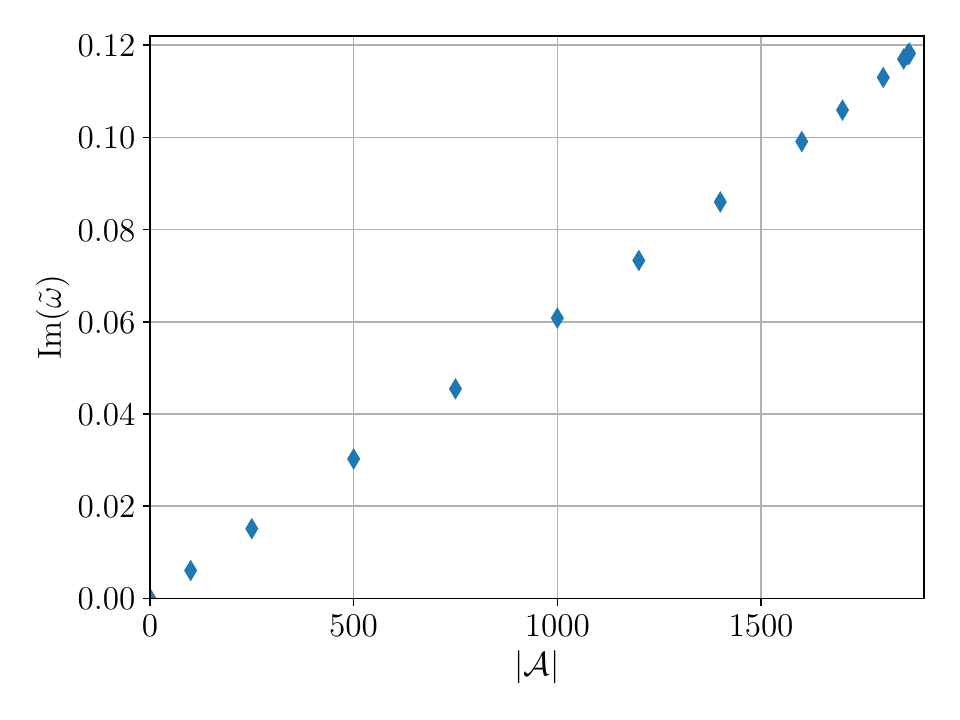}
    \caption{Plots showing how the dimensionless mode frequency of the fundamental g-mode, $g_1$, varies with $\mathcal{A}$. Plotted separately are the real and imaginary parts of the frequency on the left and right, respectively.}
    \label{ReImG1}
\end{figure}

\begin{figure}
    \centering
    \includegraphics[width=10cm]{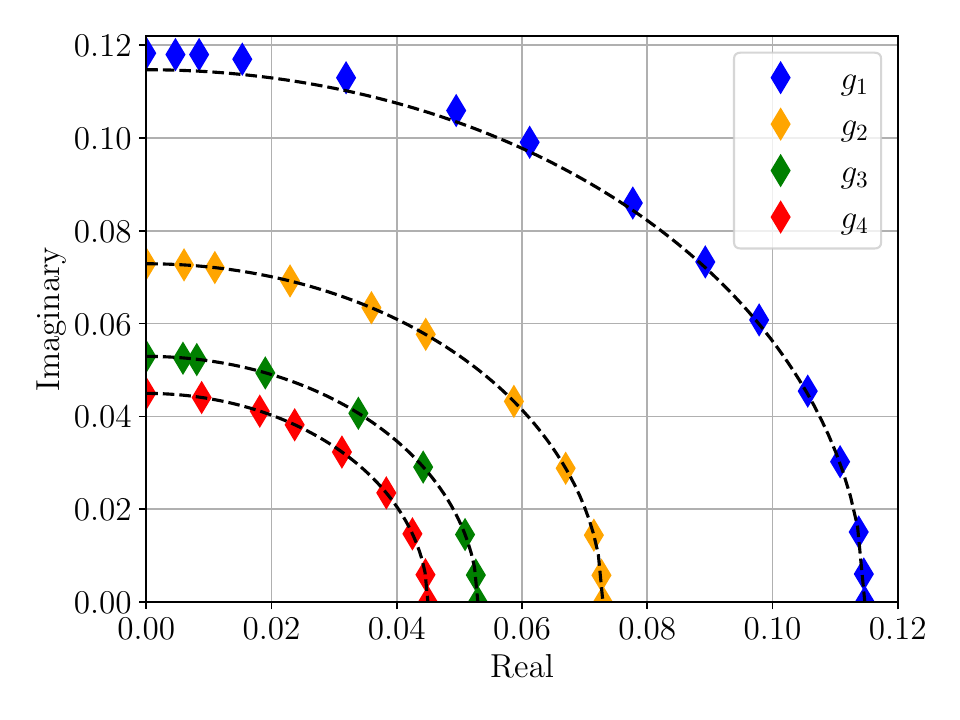}
    \caption{Results showing how the dimensionless frequencies $\tilde\omega_n$ of the first four g-modes ($g_1,g_2,g_3,g_4$) move in the complex plane when the reaction rate is varied. The diamonds represent the numerical values and the dashed black lines represent the theoretical prediction of a circle with a radius equal to the numerical value of the g-mode frequency when $\mathcal{A}\approx0$.  }
    \label{Trends}
\end{figure}

For slow reaction rates, $t_r>1$s, the modes exhibit minuscule damping, Re$(\omega) \gg$ Im$(\omega)\approx0$. The expected modes are found; the f-mode and the g- and p-modes with their respective overtones. When $t_r<1$s, the damping begins to increase for all modes. As expected, the effect is greater for the g-modes due to their relatively low frequencies in comparison to the f- and p-modes. As can be seen in Figure \ref{ReImG1}, as $|\mathcal{A}|$ increases, which corresponds to $t_r \xrightarrow{} 0$, the real part of the mode frequency decreases while the imaginary part increases. From Figure \ref{Trends} we see that the g-modes tend to move in unison with each other until a mode hits the relevant critical reaction rate, at which point the real part of the frequency quickly tends to $0$ and the mode becomes ``purely'' imaginary, Re$(\omega)\approx0$.

Beyond the critical reaction rate, 
a pair of pure imaginary modes are found for each specific g-mode, $g_n$. As the reaction rate becomes faster, the pair diverge, both staying on the imaginary axis but one increasing and the other tending towards the origin. This agrees with the behaviour seen in the plane-wave analysis of the fast reaction limit. The existence of the second mode, $\omega_2$, is due to $\omega_2=-$Re$(\omega_1)+i$Im$(\omega_1)$  being a solution if $\omega_1$ is a solution with positive real and imaginary components. The presence of two solutions is down to the actual eigenvalue being $\tilde \omega_n^2$. We can think of the two solutions physically as  modes propagating in opposite directions around the star, which due to the lack of rotation are equivalent. This obviously does not impact the mode damping, hence why the imaginary part of the frequency is the same for both solutions. 

As expected, the lowest frequency g-modes are the first to hit their critical reaction rate and reach the imaginary axis. This is akin to the ``switch off'' predicted in the previous discussion.
In agreement with the plane wave argument, we also see that the frequency at which the g-modes first hit the imaginary axis is approximately the imaginary counterpoint of the real-valued g-mode frequencies in the slow reaction limit. Sample numerical data is shown in Table \ref{Table},  showing good agreement with the predicted behaviour in the fast reaction limit. We also see that, as the g-modes sweep through the complex plane they appear to trace a circular path, with centre at the origin and radius of the initial slow-reaction limit mode frequency. The solution with a negative real part traces a symmetric pattern in the second quadrant of the complex plane. Once the g-modes hit the imaginary axis, the symmetry  breaks and the two mode solutions diverge along the imaginary axis, as previously discussed.

\begin{table}
    \large
    \centering
    \caption{In the table are the dimensionless frequencies of the first 10 g-modes, calculated numerically for $\mathcal{A}=0$ and for when the mode frequency first becomes purely imaginary which we will call the critical frequency, $\mathcal{A}_{Crit}$. The ratio is defined as $\frac{|\mathcal{A}_{Crit}|}{Im(\omega_{\mathcal{A}=0})}$. As we can see this agrees well with the predicted value of 2}
    \begin{tabular}{||c c c c c||} 
     \hline
     Mode & $\Tilde{\omega}_{\mathcal{A}=0}$  & $\Tilde{\omega}_{\mathcal{A}_{Crit}}$ &  $|\mathcal{A}_{Crit}|$& Ratio  \\ [0.5ex] 
     \hline\hline
     $g_1$ & 0.115  & 0.118i & 1864.2 & 1.84  \\ 
     \hline
     $g_2$ &  0.0729  & 0.0729i & 1264.3 & 1.97  \\
     \hline
     $g_3$ & 0.0529  & 0.0529i & 910.6 & 1.95  \\
     \hline
     $g_4$ & 0.0450  & 0.0450i & 764.9 & 1.93 \\
     \hline
     $g_5$ & 0.0380  & 0.0380i & 649.2 & 1.94  \\
     \hline
     $g_6$ & 0.0328  & 0.0328i & 560.4 & 1.94  \\
     \hline
     $g_7$ & 0.0288  & 0.0288i & 491.5 & 1.94  \\
     \hline
     $g_8$ & 0.0257  & 0.0257i & 438.1 & 1.94  \\
     \hline
     $g_9$ & 0.0232  & 0.0231i & 395.3& 1.94  \\
     \hline
     $g_{10}$ & 0.0211  & 0.0211i & 359.9 & 1.94  \\ [1ex] 
     \hline
    \end{tabular}
    \label{Table}
\end{table}

\section{Conclusions}\label{Conclusions}

Expanding on the work of \cite{10.1093/mnras/stz2449} we have shown how the presence of nuclear reactions in a neutron star leads to a damping of the composition g-modes for arbitrary reaction rates. This was achieved by setting up linear perturbation equations in a dimensionless formalism similar to that of \cite{1989nos..book.....U} and others. We showed that, as the reaction rate increases, the mode frequencies sweep through the complex plane from the real axis to the imaginary axis. As a consequence, the higher order modes are the first to be removed from the oscillation spectrum. This continues until, at a certain reaction rate, there would be no oscillatory g-modes left in the neutron star. The good agreement between our numerical results and those derived from our plane-wave analysis provides strong confidence in these conclusions. 

These results are,  however, currently mainly relevant in a phenomenological sense due to the simplifications made in the calculations, such as using Newtonian gravity. Still, we learn that we need to be careful with problems where the higher order g-modes play a role, such as the p-g instability and mode excitation during a dynamical tide, due to the impact reactions have on the mode spectrum. 

Following on from these results, the next step would be to consider more realistic background stars by extending this calculation to general relativity. One could also consider the impact temperature profiles and gradients would have on the g-modes as one of the assumptions made was that we were using a mature, cold neutron star. For example, work  by \cite{HotGmodes} shows that the g-mode frequencies and damping times can be dramatically affected by high temperature.  Finite temperatures also bring in a new class of g-modes, thermal g-modes, in addition to the composition modes we have considered here. This future direction would  open the way to considering objects like proto-neutron stars formed after the gravitational collapse of successful core–collapse supernova explosions \citep{PNS1,Burrows}, systems where nuclear reactions will also have an impact on the dynamics.

\section*{Acknowledgements}

NA acknowledges support from STFC via grant number ST/R00045X/1.


\section*{Data Availability}

Additional data related to this article will be shared on reasonable request to the corresponding author.

\bibliographystyle{mnras}
\bibliography{Gmodebibliography}

\end{document}